# Comoving acceleration of overdense electron-positron plasma by colliding ultra-intense laser pulses

Edison Liang[1]

[1]*Rice University*

Abstract

Particle-in-cell (PIC) simulation results of sustained acceleration of electron-positron (e+e-) plasmas by comoving electromagnetic (EM) pulses are presented. When a thin slab of overdense e+e- plasma is irradiated with linear-polarized ultra-intense short laser pulses from both sides, the pulses are transmitted when the plasma is compressed to thinner than ~ 2 relativistic skin depths. A fraction of the plasma is then captured and efficiently accelerated by self-induced **J**x**B** forces. For 1μm laser and $10^{21}$Wcm$^{-2}$ intensity, the maximum energy exceeds GeV in a picosecond.

Recent advances in ultra-intense short-pulse lasers (ULs) [1,2] revolutionize particle acceleration via intense electromagnetic (EM) fields [3]. Most proposed laser acceleration schemes (e.g. laser wakefield accelerator (LWFA), plasma wakefield accelerator (PWFA) [4], free-wave accelerator (FWA) [5]) require propagating lasers in an underdense plasma ($\omega_{pe}=(4\pi ne^2/m)^{1/2}<\omega_o=2\pi c/\lambda$, $\lambda$=laser wavelength, n=electron density, m=electron mass). In such schemes the energy gain/distance [4] and particle beam intensity are constrained by the underdense requirement. Here we report particle-in-cell (PIC) simulation results of a radically different concept: comoving acceleration of overdense ($\omega_{pe}>\omega_o$) plasmas using colliding UL pulses. This colliding laser pulses accelerator (CLPA) has properties, such as higher acceleration gradient and particle beam intensity, that are complementary to underdense schemes.

When an intense EM pulse with $\Omega_e$ (=$eB_o/mc=a_o\omega_o$, $a_o$=normalized vector potential)>$\omega_{pe}$ initially imbedded in an overdense plasma tries to escape, it induces a skin current **J** that inhibits the EM field from leaving. The induced **J x B** (ponderomotive) force then accelerates the surface plasma to follow the EM pulse. As the EM pulse "pulls" the surface plasma, it is slowed by plasma loading (group velocity < c), allowing the fastest particles to "comove" with the EM pulse. Since slower particles eventually fall behind, the plasma loading decreases and the EM pulse accelerates over time. A dwindling number of fast particles gets accelerated indefinitely by the comoving EM force, reaching maximum Lorentz factors $\gamma_{max}>a_o^2/2$ (ponderomotive limit [6]) >>$(\Omega_e/\omega_{pe})^2$. This phenomenon, called the diamagnetic relativistic pulse accelerator (DRPA) [7], is a nonlinear relativistic phenomenon, with no analog in the weak field ($\Omega_e/\omega_{pe}<1$), low density ($\omega_o>\omega_{pe}$) regime or test particle limit.

But DRPA is difficult to achieve in the laboratory, since vacuum EM waves cannot penetrate an overdense plasma beyond the relativistic skin depth [8]. Fig.1 shows the PIC simulation of a single UL irradiating an overdense e+e- plasma. All upstream plasma is snowplowed by the UL light pressure, and the asymptotic Lorentz factor stays below $\gamma_{max} \sim 45$. The relativistic snowplow compresses the density, so the plasma stays overdense and ahead of the EM pulse, preventing the UL from passing through. Even when the initial slab thickness is less than the relativistic skin depth, we find that ponderomotive snowplowing supersedes wave transmission, and the EM pulse fails to overtake the plasma (Fig.1). Hence the DRPA initial condition cannot be achieved using a single UL pulse. Using PIC simulations with the 2.5D (2D-space, 3 momenta) ZOHAR code [9], here we report that sustained comoving acceleration similar to DRPA can be achieved by irradiating a thin slab of overdense e+e- plasma with UL pulses *from both sides*. The opposing UL pulses first compress the overdense plasma to a total thickness < 2 relativistic skin depths [8] while keeping the central plasma in place. At that point the UL pulses "tunnel" through the overdense plasma ($\omega_{pe} > <\gamma>^{1/2}\omega_o$, $<\gamma>$=mean Lorentz factor of the compressed plasma). The subsequent acceleration via comoving **J**x**B** forces resembles DRPA [7].

Fig.2 shows the evolution of two linearly polarized half-cycle plane EM pulses with parallel **B**, irradiating a thin e+e- slab from opposite sides (thickness=$\lambda/2$, initial density $n_o=15n_{cr}$(critical density)). Cases with nonparallel **B** are more complex and will be reported separately. The incident pulses initially snowplow the plasma inward as in Fig.1 (Fig.2a). Only ~10% of the incident amplitudes is reflected during the compression as the laser reflection fronts move inward relativistically [10]. When the skin currents from both sides merge (Fig.2b), the two UL pulses interpenetrate and tunnel through the plasma, despite $\omega_{pe} > <\gamma>^{1/2}\omega_o$. Such overdense transmission of EM pulses occurs only because the plasma thickness is < 2 relativistic skin depths and the central plasma is *kept in place by the opposing light pressure*. As the transmitted UL pulses reemerge from the plasma, they induce new drift currents **J** at the *trailing* edge of the pulses (Fig.2c), with signs opposite to the initial currents (Fig.2b), so that the **J** x **B** forces pull the plasma outward and load the EM pulses. This plasma loading plays a crucial role in sustaining the comoving acceleration: a larger fraction of plasma is picked up and accelerated for higher initial densities (cf.Fig.4a below). As slower particles eventually fall behind the UL pulses, the UL plasma loading decreases with time. This leads to continuous acceleration of both the UL pulses and the fastest particles. The growth of $p_x$ vs. x (Fig.2d) of CLPA resembles that of DRPA [7] at late times.

Fig.3 shows the results of irradiating an overdense e+e- slab using Gaussian pulse trains ($\lambda$=1μm, pulse length τ=85fs, $I_{peak}$=$10^{21}$Wcm$^{-2}$). Here the compressed plasma slab cleanly separates into left and right-moving pulses only after the peaks of the two wave trains have passed each other. Fig.3b shows that $\gamma_{max}$ grows as a power-law in time ($\gamma_{max}$~$t^{0.8}$), reaching 2200 (1.1GeV) in 1.3ps, far exceeding the nominal ponderomotive limit $a_o^2$/2 (=360; this limits does not apply because the *instantaneous* $a_o$ increases with time due to pulse stretching, Fig.3c). We can derive this $\gamma_{max}$ growth rate from the Lorentz equation: d$\gamma_{max}$(t)/dt~eE(t)mc where E(t) is the UL electric field comoving with the fastest particles. From the output we confirm that E(t)~$t^{-0.2}$ due to energy transfer to the particles (Fig.3c shows decay of the B profile). The asymptotic particle spectrum also forms a power-law with slope ~ –1 (Fig.3d). Such power-law spectrum is distinct from the usual exponential spectrum produced by ponderomotive stochastic heating [11,12]. A power-law is formed since there is no other preferred energy scale below $\gamma_{max}$, and the particles develop random phases over time with respect to the EM field profile. In practice, $\gamma_{max}$ is limited by the diameter D of the laser focal spot, since particles drift transversely out of the laser beam after t~D/c. Dimensionally, the maximum energy of any comoving acceleration is thus <e$E_o$D=6GeV(I/$10^{21}$Wcm$^{-2}$)$^{1/2}$(D/100μm). We find no evidence of any transverse instability, which is suppressed by relativistic effects plus strong transverse **E, B** fields that oppose momentum isotropization.

Fig.4 compares results of different laser and plasma parameters. Fig4a shows that both spectral hardness and $\gamma_{max}$ increase with intensity, while particle pickup increases with plasma density. In Fig4b we show variation with pulse length $\tau$. At first $\gamma_{max}$ increases but the power-law slope stays constant as we increase $\tau$. But for long pulses, $\gamma_{max}$ ~constant while the slope hardens with increasing $\tau$. Fig.4c compares the energy coupling efficiency for sample intensities and densities. We find that EM energy coupling to particles increases with intensity and with density, reaching a maximum of 45% among the runs completed so far. After saturation the particle and EM energies oscillate (curves D–G) because, while the fastest particles continue to accelerate, the slower tail particles transfer energy back to EM waves at late times. Fig.4d compares the energy-angle distributions for sample intensities and densities. The highest energy particles are narrowly beamed. We have scratched only the tip of the vast CLPA parameter space. A full exploration is underway but will require many years of systematic studies.

An experimental demonstration of the CLPA will require a dense and intense e+e- source. Cowan et al [13] demonstrated that such an e+e- source can be achieved using a PW laser striking a gold foil. Theoretical works [14] suggest that e+e- densities $>10^{22}cm^{-3}$ may be achievable with sufficient laser fluence. Such a dense e+e- jet can be slit-collimated to produce a < micron thin e+e- slab, followed by 2-sided irradiation with opposing UL pulses. For example, UL pulses with $\tau=80fs$, peak intensity=$10^{21}Wcm^{-2}$ and spot diameter D=15μm require 1.8PW peak power and 70J, within the range of UL's currently under construction. Fig.3bd (black curves) suggest that pairs can be accelerated to a power-law with $E_{max}>100MeV$, easily distinguishable from an exponential spectrum with kT~16MeV produced by ponderomotive heating [12]. Note that if one pulse arrives first, it simply pushes the plasma until the opposing pulse hits. The subsequent evolution is similar to the simultaneous arrival cases reported above.

We have also studied the effects of finite laser spot size. Preliminary results suggest that for sufficiently uniform core intensity, our plane wave results remain valid in the core.

**ACKNOWLEDGEMENTS**

EL was supported by NASA NAG5-7980, LLNL B537641 and NSF AST-0406882. He thanks Scott Wilks for helps with ZOHAR and useful comments, and Bruce Langdon for providing the ZOHAR code.

Figure Captions

FIG.1 PIC simulation shows that a single UL pulse ($I(\lambda/\mu m)^2=10^{21}W/cm^2$, $c\tau=\lambda/2$) snowplows an overdense ($n_o=15n_{cr}$, thickness = $\lambda/2$, $kT=2.6keV$) e+e- plasma indefinitely, but cannot pass through to load the plasma on the backside. We plot $B_y$ (in units of $0.8mc\omega_o/e$), $n/n_{cr}$ and $p_x/mc$ (black dots) vs. $16x/\lambda$ at $t\omega_o = 40\pi$. The maximum Lorentz factor $\gamma_{max}$ remains $\leq 45$ in this case.

FIG.2 Evolution of two linearly polarized plane EM pulses ($I(\lambda/\mu m)^2=10^{21}W/cm^2$, $c\tau=\lambda/2$) irradiating an overdense e+e- plasma slab centered at $8x/\lambda=180$ ($n_o=15n_{cr}$, thickness = $\lambda/2$, $kT=2.6keV$) from opposite sides. We plot magnetic field $B_y$, electric field $E_z$ (in units of $0.8mc\omega_o/e$) current density $J_z$ (in units of $0.05mc\omega_o^2/e$) and $p_x/mc$ vs. $8x/\lambda$ (inset) at $t\omega_o/2\pi =$ (a)1.25, (b)1.5, (c)1.75; (d) Snapshots of $p_x/m_ec$ vs. $8x/\lambda$ for the right-moving pulse at $t\omega_o/2\pi=$(left to right) 2.5, 5, 10, 22.5, showing continuous growth of $\gamma_{max}$. We also show the profiles of $B_y$, $E_z$ (same units as above) at $t\omega_o/2\pi=22.5$.

FIG.3 (color) Results of two Gaussian pulse trains ($\lambda=1\mu m$, $I=10^{21}W/cm^2$, $c\tau=85fs$) irradiating an e+e- plasma centered at $2\pi x/\lambda=4800$ ($n_o=9n_{cr}$, thickness = $2\lambda/\pi$, $kT=2.6keV$) from both sides. (a) $B_y$ (in units of $mc\omega_o/e$) and $n_o/n_{cr}$ profiles at $t\omega_o=0$; (b) plot of $\log(p_x/m_ec)$ vs. $\log(2\pi x/\lambda -4800)$ for the right-moving pulse at $t\omega_o=$180(black), 400(green), 800(cyan), 1600(yellow), 2400(magenta), 4000(blue), 4800(red) showing power-law growth of $\gamma_{max}\sim t^{0.8}$; (c) detailed profiles of the left-moving pulse at $t\omega_o=4800$: we plot $p_x/10000$ (blue dots), $B_y/100$ (black, same units as above), $n/n_{cr}$ (red) vs. $2\pi x/\lambda$; note the back-half of the UL pulse has mostly decayed due to energy transferred to the particles; (d) evolution of electron spectrum $f(\gamma)$ (normalized distribution of particles per unit $\gamma$) vs. $\gamma$ showing the build-up of power-law with slope approaching -1 (lower solid line): $t\omega_o=$180(black), 400(green), 800(cyan), 2400(magenta), 4800(red).

FIG.4 (color) (a) Normalized electron spectrum at $t\omega_o=4800$ of Fig.3 (red) compared to that with $I=4\times10^{19}$W/cm$^2$ (black) and $n_o/n_{cr}=0.4$ (blue), other parameters the same as in Fig.3. (b) Normalized electron spectra at $t\omega_o/2\pi=22.5$ when we vary the pulse length $c\tau= \lambda/2$(black), $\lambda$(green), $4\lambda$(blue) $7\lambda$(magenta), $26\lambda$(red), other parameters the same as in Fig.2. (c) Plots of particle energy (blue), field energy (red) and total energy (C, black) with time. We compare the case of Fig.3 (D, E) with $I=4\times10^{19}$W/cm$^2$ (A, B) and $n_o/n_{cr}=25$ (F, G), other parameters the same as in Fig.3. (d) Comparison of $\gamma$ vs. angle (from x-axis) in x-z plane for the case of Fig.3 (red) with $I=4\times10^{19}$W/cm$^2$ (black), $I=1.1\times10^{22}$W/cm$^2$ (blue) and $n_o/n_{cr}=25$ (green), other parameters the same as in Fig.3. Higher density and lower intensity lead to slightly broader angular distribution.

Fig.1

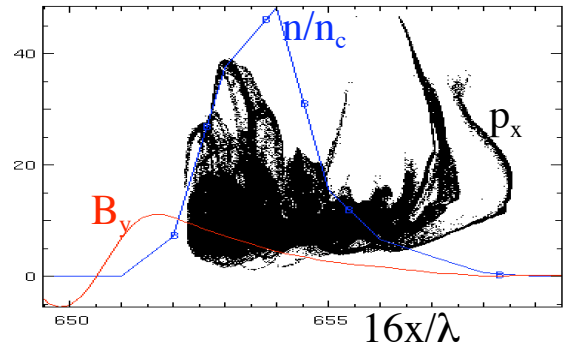

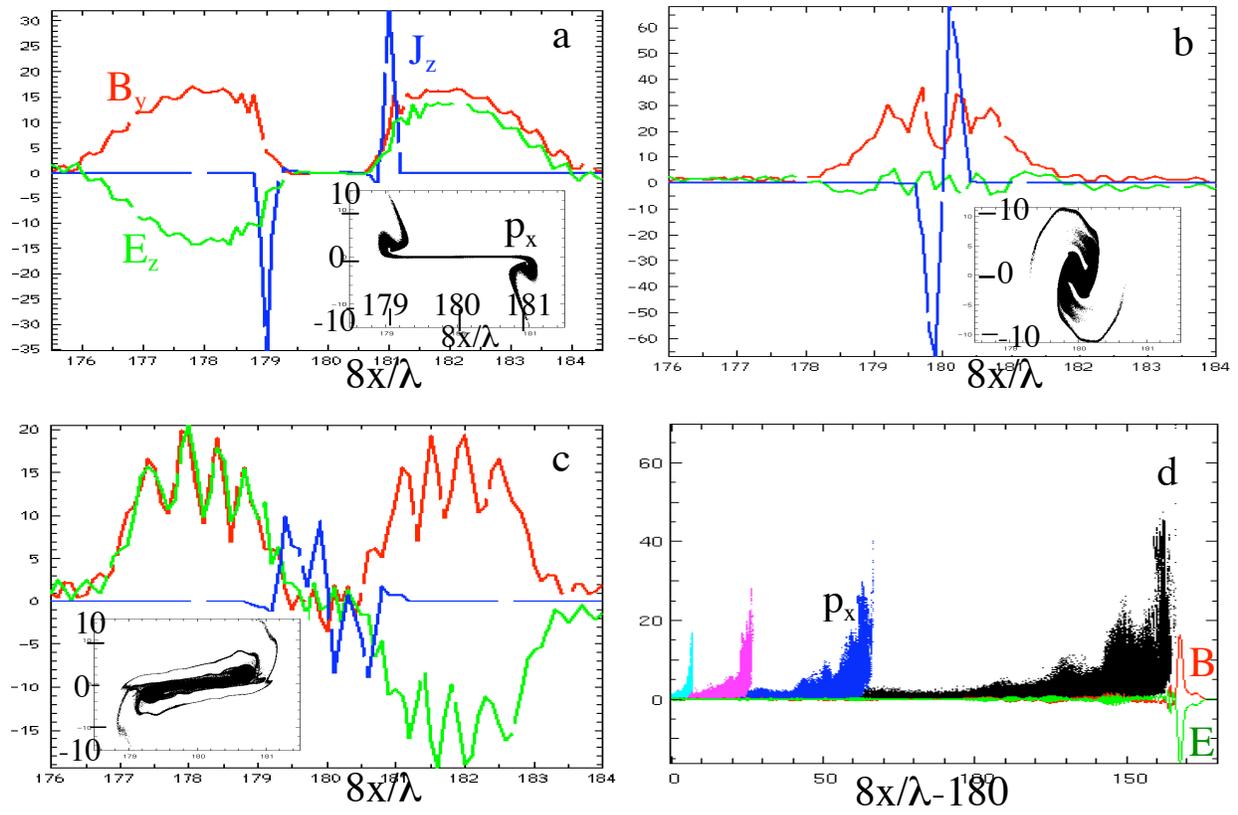

Fig.2

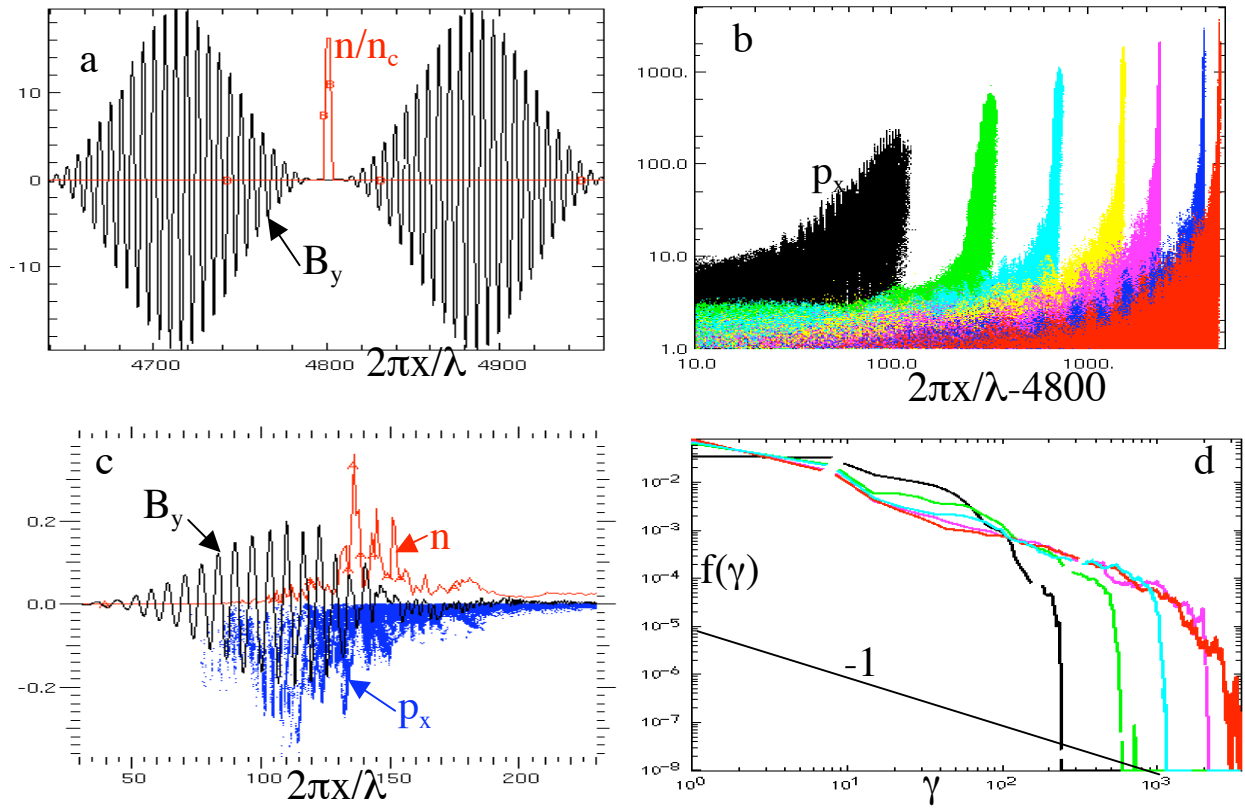

Fig.3

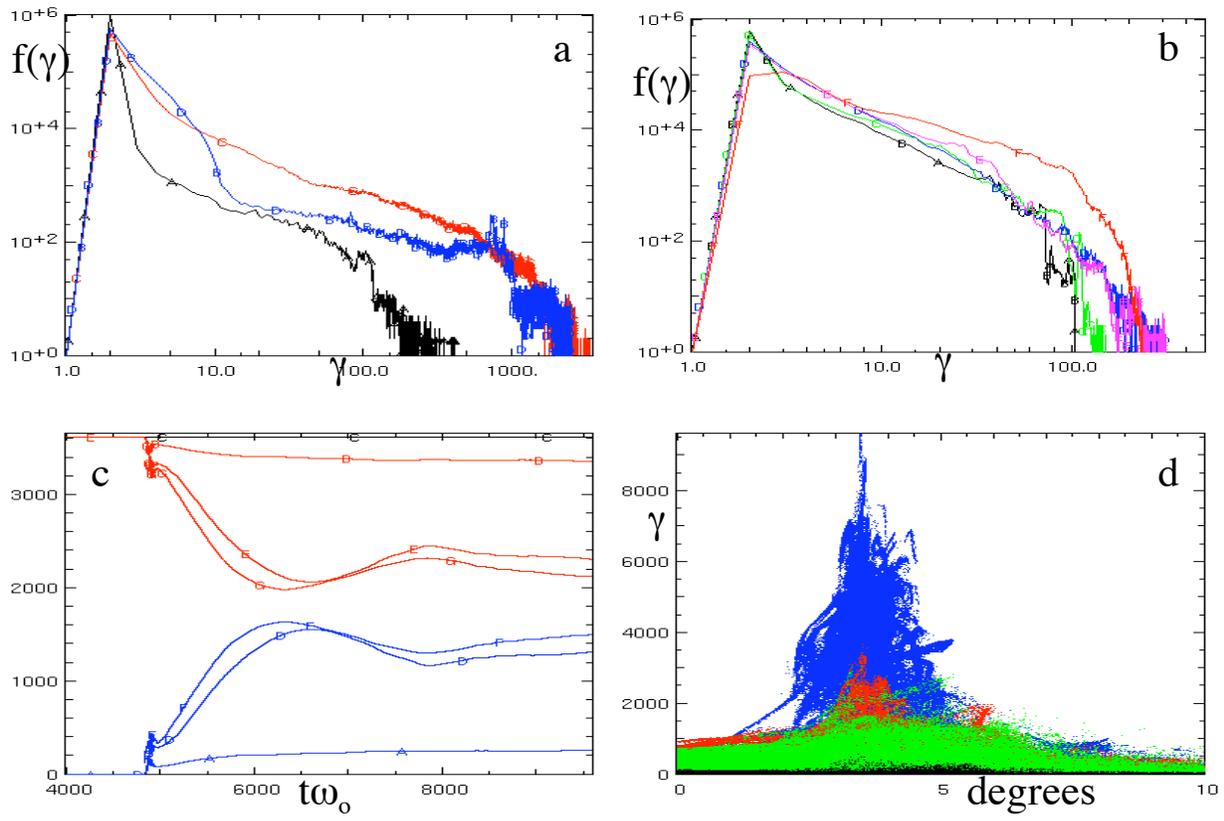

Fig.4